\documentclass[12pt]{iopart}
\usepackage{bm}
\usepackage{graphicx}
\usepackage{iopams}
\begin{document}
\title{Atomic swelling upon compression}
\author{ V K Dolmatov and J L King\footnote{Presently a graduate student at the Department of Physics, Auburn University, Auburn, Alabama 36849, U.S.A.}}
\address{Department of Physics and Earth Science, University of North Alabama,
Florence, Alabama 35632, USA}
\ead{vkdolmatov@una.edu}
\begin{abstract}
The hydrogen atom under the pressure of a spherical penetrable confinement potential of a decreasing radius $r_{0}$ is explored, as a case study. A novel
counter-intuitive effect of atomic swelling rather than shrinking with decreasing
 $r_{0}$ is unraveled, when $r_{0}$ reaches, and remains smaller than, a certain critical value. Upon swelling, the size of the atom is shown to increase
by an order of magnitude, or more, compared to the size of the free atom. Examples of changes of photoabsorption properties of  confined hydrogen atom upon
its swelling are uncovered and demonstrated.
\end{abstract}
\submitto{\jpb}
\maketitle
Modifications in the structure and spectra of atoms confined in cages whose sizes are commensurable with atomic sizes has been probed by researchers
since the early works of Michels \textit{et al} \cite{Michels} and
Sommerfeld and Welker \cite{Sommerfeld} devoted to hydrogen centrally confined in an impenetrable square-well of adjustable radius, to simulate pressure. To date,
numerous aspects of the structure and spectra of atoms under various kinds of confinements have been attacked from many different angles by
research teams world-wide. This has resulted in a huge array of unraveled effects and data being accumulated in a large number of publications, see reviews
\cite{Jaskolski,Buchachenko,RPC'04} as well as numerous review papers in \cite{AQC57,AQC58} (and references therein). There, one finds a wealth of information
on properties of single-electron, two-electron
and many-electron atoms confined  by impenetrable spherical, spheroidal as well as open boundaries potentials
 (e.g., see review papers in \cite{AQC57} by Aquino, p.$123$; Laughlin, p.$203$; Cruz, p.$255$;   Garza and Vargas, p.$241$),
oscillator potentials (e.g.,  Patil and Varshni \cite{AQC57}, p.1), potentials limited by conoidal boundaries (Ley-Koo E \cite{AQC57}, p.79), Debye potentials
(Sil, Canuto and Mukherjee \cite{AQC58}),
fullerene-cage potentials (Dolmatov \cite{AQC58}, p.13; Charkin \textit{et al} \cite{AQC58}, p.69), \textit{etc}. All these activities speak to the importance of the subject.
This is because  confined atoms behave differently from free atoms in ways which provide
insight into various interesting problems of interdisciplinary importance. To list a few, the latter could be associated with atoms trapped in hollow cavities of solids, zeolites,
fullerenes, helium droplets formed in walls of nuclear reactors, atoms/ions placed in a plasma environment, \textit{etc}. Furthermore, by suitably tailoring the
confinement parameters, novel specific atomic properties can be designed in a controllable manner, thereby opening up
  new technological possibilities for confined atoms.

The aim of this paper is to demonstrate that upon compression of  an atom by a suitably tailored repulsive confining spherical potential $U_{\rm c}(r)$ of a finite height and thickness, the atom can be transformed into novel exotic sort of
an atom of a large size and distinct spectra. It is found that upon decreasing the radius of the confining potential (thereby increasing the pressure on a confined atom) below to a certain critical value $r_{\rm c}$, the atom stops
behaving in the conventional manner. Instead, the atom suddenly swells rather than keeps shrinking in size. This effect is termed \textit{atomic swelling}.
 It is the ultimate aim of this paper to demonstrate and interpret atomic swelling, as well as to explore trends which might occur in photoabsorption spectra of confined atoms upon atomic swelling. The hydrogen atom is chosen as a touchstone for such a study. Atomic units (a.u.) are used throughout the paper unless otherwise specified.

For confined hydrogen, radial wavefunctions $P_{n(\epsilon)\ell}(r)$ and energies $E_{n(\epsilon)\ell}$ of discrete states $n\ell$ or continuum spectrum $\epsilon\ell$, in the presence of
 a spherical confinement modeled by a confining potential $U_{\rm c}(r)$, are determined by a radial Schr\"{o}dinger equation
\begin{eqnarray}
\fl -\frac{1}{2}\frac{d^2P_{n(\epsilon) \ell}}{dr^2} +\left [\frac{-1}{r} +\frac{\ell(\ell+1)}{2 r^2} +U_{\rm c}(r)\right ]P_{n(\epsilon)\ell}(r) \nonumber \\
\fl = E_{n(\epsilon) \ell} P_{n(\epsilon)\ell}(r).
\label{EqH}
\end{eqnarray}
As a first guiding step in understanding which aspects of a confined/compressed hydrogen atom (labelled H@$U_{\rm c}$) are most unusual,
$U_{\rm c}(r)$ is approximated by a square-well potential of certain adjustable
inner radius $r_{0}$, height $U_{0}>0$ and thickness $\Delta$, as in \cite{ConDolMan00}:
\begin{eqnarray}
U_{\rm c}(r)=\left\{\matrix {
U_{0}>0, & \mbox{if $r_{0} \le r \le r_{0}+\Delta$} \nonumber \\
0 & \mbox{otherwise.} } \right.
\label{SWP}
\end{eqnarray}
However, the square-well with infinitely sharp boundaries
might be thought to induce some artefacts in the structure and spectra of H@$U_{\rm c}(r)$. Therefore, a trial study where the square-well potential is replaced by a potential with diffuse boundaries
[to be labelled as $U_{\rm c}^{\rm DP}(r)$] is
conducted in this paper as well. For this, $U_{c}^{\rm DP}(r)$ is represented by the sum of two repulsive Woods-Saxon potentials:
\begin{eqnarray}
\fl U_{\rm c}^{\rm DP}(r) & = \left. \frac{2U_{0}}{1+{\exp}(\frac{r_{0} -r}{\eta})}\right |_{r \le r_{0} +\frac{1}{2}\Delta} \nonumber\\
\fl & +\left. \frac{2U_{0}}{1+{\exp}(\frac{r-r_{0}-\Delta}{\eta})}\right |_{r >r_{0}+\frac{1}{2}\Delta}.
\label{DP}
\end{eqnarray}
Here, $\eta$ is the diffuseness parameter, and $r_{0}$, $U_{0}$ and $\Delta$ are the same as the parameters of the square-well potential (\ref{SWP}).
In this paper, the $U_{0}$, $\Delta$ and $\eta$ parameter values are arbitrary chosen to be  $U_{0}=2.5$, $\Delta = 5$  and $\eta=0.5$, just as a case study.

One of key results of the study - atomic swelling upon compression - is demonstrated by figure $1$ for the $1\rm s$ ground-state of hydrogen under  compression.
\begin{figure}[h]
\center{\includegraphics[width=8cm]{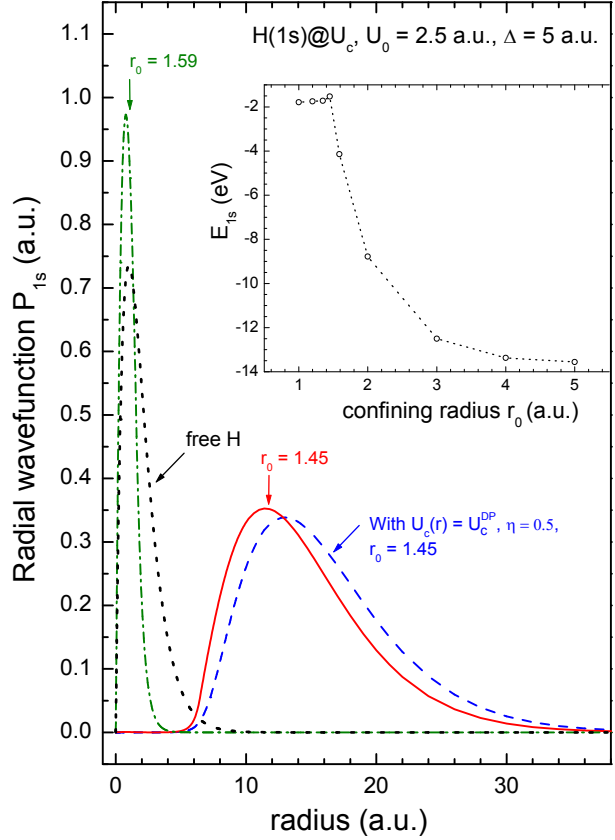}}
\caption{ The $1\rm s$ radial wavefunction $P_{1s}(r)$ of hydrogen confined by a square-well potential
with $U_{0}=2.5$ a.u. and $\Delta = 5$ a.u. calculated for $r_{0} =5$, $1.59$ and $1.45$ a.u., as marked. Also plotted are calculated data
obtained with the use of a diffuse confining potential $U^{\rm DP}(r)$ with $\eta = 0.5$, as well as data for free hydrogen, as marked. Inset: The
$1\rm s$ electron energy versus the confining radius $r_{0}$ of the square-well potential.}
\label{fig1}
\end{figure}
First, one can see that the  $P_{1\rm s}(r)$ wavefunction of H@$U_{\rm c}$ gets contracted into the inner region of space
as $r_{0}$ is decreased from $r_{0}=\infty$ (free hydrogen) to  $r_{0}=1.59$. The energy $E_{1\rm s}$, in turn,
 increased (less binding) from $E_{1\rm s}\approx -13.6$ eV for free hydrogen to
 $E_{1\rm s}\approx -4.14$ eV for H@$U_{\rm c}$ at $1.59$. Thus far, both $P_{1\rm s}(r)$ and $E_{1\rm s}$
 behave in the conventional manner. However, at just a bit smaller $r_{0}$, specifically, at $r_{0}= 1.45$  versus $r_{0}= 1.59$, the $1\rm s$ orbital suddenly expands
into an outer region in space, gets quite diffuse and peaks at $r \approx 11.5$   or  $r \approx 13$, depending on whether the atom is confined by the square-well or diffuse potential, respectively.
The implication is that under the increased pressure, when $r_{0}$ reaches the value of $r_{0}=1.45$, the atom suddenly swells strongly, rather than getting shrunk, in size - the effect
referred to as \textit{atomic swelling} in this paper. At this $r_{0}=1.45$, the atomic size becomes more than by the order of magnitude bigger that the size of the free atom,
due to spectacular atomic swelling.
A trial calculation showed that further decrease of $r_{0}$ hardly affects $P_{1\rm s}(r)$ and $E_{1\rm s}$; both of them
remain practically unchanged at $r_{0} \le 1.45$. This is because, at $r_{0} \le 1.45$,  about all of the $1\rm s$ electron density has concentrated outside of the
confining potential. Therefore, decreasing $r_{0}$ below $1.45$ cannot exert any more pressure on the atom. Hence, both $P_{1\rm s}(r)$ and $E_{1\rm s}$ stop changing any significantly for all $r_{0} \le 1.45$.
 Next, the $P_{1\rm s}(r)$ functions, calculated with the use of either the square-well or diffuse confining potential,
are almost the same at $r_{0}=1.45$, as is clearly seen in figure $1$. This implies that, first, atomic swelling is not an artifact caused by the
infinitely sharp boundaries of the square-well potential
and, second, atomic swelling is somewhat insensitive to the shape of a penetrable confining potential. For this reason, all other calculated data  presented in this paper were obtained
 with the use of the square-well potential, for the sake of simplicity. Interestingly, this effect of atomic swelling is opposite to another
 counter-intuitive effect of \textit{orbital compression} by \textit{attractive confinement} which was revealed earlier in work \cite{ConDolMan99}. Finally,
 note, the attempt to calculate $P_{1\rm s}(r)$ and $E_{1\rm s}$ between $1.45 < r_{0} < 1.59$
failed, for a reason which is explained later in the paper.

The physics behind the effect discovered, atomic swelling of hydrogen under compression, becomes clear when one explores  figure $2$, where the effective potential  $U_{1\rm s}^{\rm eff}(r) = -\frac{1}{r} + U_{\rm c}(r)$
`seen' by the $1\rm s$ electron in the H@$U_{\rm c}$ atom is depicted.
\begin{figure}[h]
\center{\includegraphics[width=8cm]{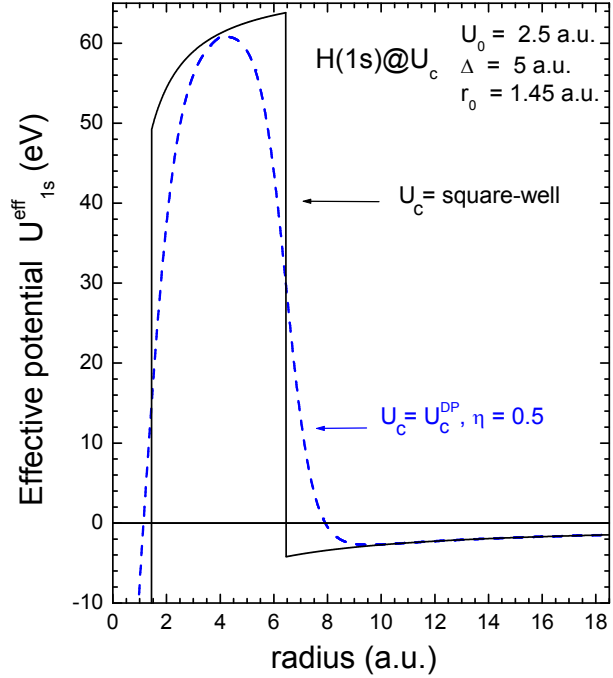}}
\caption{The effective potential  $U_{1\rm s}^{\rm eff}(r) = -\frac{1}{r} + U_{\rm c}(r)$
``seen'' by the $1\rm s$ electron in the H@$U_{\rm c}$ atom when the confining potential $U_{\rm c}$ is approximated by a square-well potential with $U_{0} = 2.5$, $r_{0}= 1.45$ and $\Delta = 5$ a.u. (solid line),
    or diffuse potential $U_{\rm c}^{\rm DP}$ with $\eta = 0.5$, as marked.}
\label{fig2}
\end{figure}
One can see that adding the confining potential to the atomic potential makes $U_{1\rm s}^{\rm eff}(r)$ to consists of two wells, a short-range inner well and shallow long-range outer well. As long as the
the confining radius $r_{0}$ is such that the inner short-range is more binding than the outer well, the $1\rm s$ electrons remains in the inner well. There, the atom behaves `normally', i.e.,
its size is shrinking and the $E_{1\rm s}$ energy rising with decreasing $r_{0}$. However, as the confinement radius $r_{0}$ and, hence, the width of the inner short-range well becomes the same or smaller than some critical value,
 $r_{0} \le  r_{\rm c}$, the inner well becomes less binding, and so the binding of the electron is altered in favour of the long-range outer well. As a result, atomic swelling into the outer well is induced by gradually
reducing the confinement radius $r_{0}$ to $r_{0} \le r_{\rm c}$. In the chosen case study, atomic swelling occurred at $r_{0} \le 1.45$. Note, the double-well potential naturally occurs in d- and f-series of free atoms,
resulting in the effect know as orbital collapse (a thorough review of the topic was given by Connerade \cite{Connerade_book}, Chapter $5$). The difference here is that orbital collapse is due to the cetrifugal term in the effective atomic potential, and, hence, affects only states with $\ell\neq 0$, in contrast to the present study. It also results in orbital shrinking rather than swelling. Both situations, however, are clearly similar in the spirit. Furthermore, even closer in the spirit, but opposite in the intention and outcome, was the study performed in \cite{ConDol,ConDolLaksh}. There, atoms with an electron being originally bound by an outer long-range well of the \textit{natural} atomic double-well potential [e.g., an excited $3\rm d^*$ electron in Cr($\rm 3p^{5}3d^{5}4s^{1}3d^*$, $^{7}\rm P$)], were placed inside a spherical potential of decreasing radius $r_{0}$, with finite or infinite potential height $U_{0}$,
but infinite thickness $\Delta \rightarrow \infty$. That resulted in turning the outer long-range well into a  short-range well which, thus, became less binding, or otherwise destroyed at all, at small $r_{0}$, for an obvious reason. As a result, the competition
between the inner and outer wells was altered in favour of the former, leading to $3\rm d^*$ orbital collapse into the inner well. In the present study, on the contrary,
due to \textit{finite} $\Delta$, the double-well potential is artificially \textit{created} rather than getting destroyed as in \cite{ConDol,ConDolLaksh}, and not the outer well  but the \textit{inner} well is shortened upon decreasing $r_{0}$, so that the competition takes the opposite direction resulting in orbital swelling rather than collapse.

Now, what about the range of $1.45 < r_{0} < 1.59$, where the calculation of $P_{1\rm s}(r)$ and $E_{1\rm s}$ failed in the present study? The interpretation is that, for  $1.45 < r_{0} < 1.59$,
both the short-range inner and long-range outer wells have about the same binding strength for the $1\rm s$ electron. Correspondingly,  $P_{1\rm s}(r)$ should possess two maxima of not too dissimilar amplitudes,
one in each of the well. Earlier such rare occurrence was found to emerge naturally in  $n\rm f$ series of Ba$^{+}$ \cite{Connerade_book,ConMansf} and in compressed Cr \cite{ConDol,ConDolLaksh}. It becomes very difficult to obtain the solution of the Shr\"{o}dinger  or Hartree-Fock equation in this parameter range, where a standard computer algorithm becomes unstable; see  discussion of this topic in Chapter $5$ of \cite{Connerade_book}, or a brief discussion and references in \cite{ConDol}.

Not only can the $1\rm s$ ground state of confined hydrogen experience atomic swelling, but excited states as well. This is clearly demonstrated by figures $3$ and $4$, where the $P_{2\rm p}$ and $P_{3\rm p}$ radial functions  of hydrogen compressed by the square-well potential with  $r_{0} =1. 45$ (figure $3$) and $r_{0}=1.59$ (figure $4$)  are depicted.
\begin{figure}[h]
\center{\includegraphics[width=8cm]{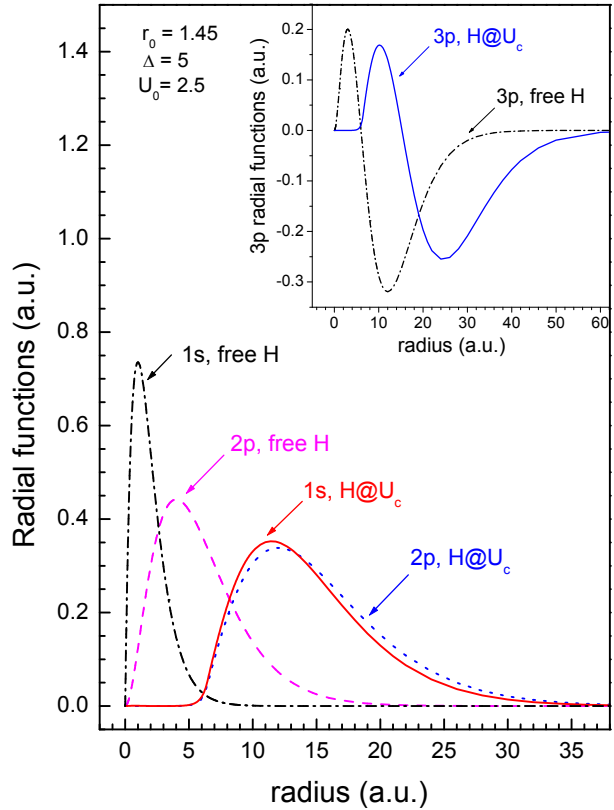}}
\caption{Radial functions $P_{1\rm s}(r)$, $P_{2\rm p}(r)$ and $P_{3\rm p}(r)$ of free hydrogen and hydrogen confined by the square-well potential $U_{\rm c}$ with $U_{0} = 2.5$, $r_{0}= 1.45$ and $\Delta = 5$ a.u., as marked.}
\label{fig3}
\end{figure}
\begin{figure}[h]
\center{\includegraphics[width=8cm]{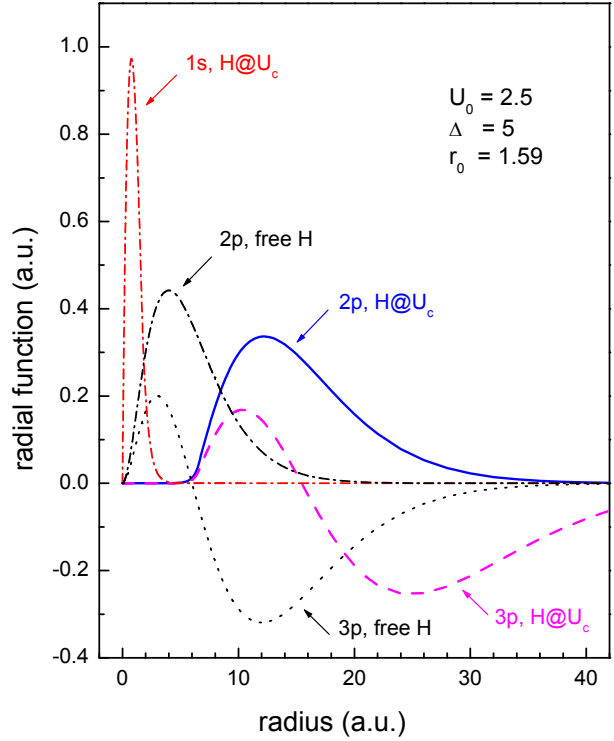}}
\caption{Radial functions $P_{1\rm s}(r)$, $P_{2\rm p}(r)$ and $P_{3\rm p}(r)$ of free hydrogen and hydrogen confined by the square-well potential $U_{\rm c}$ with $U_{0} = 2.5$, $r_{0}= 1.45$ and $\Delta = 5$ a.u., as marked.}
\label{fig4}
\end{figure}
Let us discuss the displayed calculated data for $r_{0}=1.45$ and $r_{0}=1.59$ separately, along with their significance for the photoabsorption spectra of the compressed atom.

Exploring figure~$3$ ($r_{0}=1.45$), one can see, first, that the excited $P_{2\rm p}$ orbital peaks at $r \approx 12$ versus $r \approx 4$  for free hydrogen. Thus, the size of the   H($2\rm p$)@$U_{\rm c}$
 atom is three times of the size of the excited free atom -  a clear evidence of atomic swelling of excited states under the confinement.
Second, highly spectacular, the excited $P_{2\rm p}$  and ground-state $P_{1\rm s}$ functions of H@$U_{\rm c}$ appear to be about the same and pick at
about the same value of $r$. The overlap between $P_{2\rm p}$  and  $P_{1\rm s}$  is thus huge, compared to that of the free atom. This implies an anomaly large oscillator strength $f_{1\rm s \rightarrow 2p}$
of the ${1\rm s} \rightarrow {2\rm p}$ transition in H@$U_{\rm c}$. Indeed, the calculation shows that $f_{1\rm s \rightarrow 2p} \approx 0.812$ in H@$U_{\rm c}$ compared to only $\approx 0.416$ in free hydrogen.
As known, the sum of all oscillator strengths of $n\ell \rightarrow n'(\epsilon)\ell\pm1$ transitions from an atomic $n\ell$ subshell into its discrete and continuum spectra
 equals the number of electrons in the subshell. Hence, in our case, over $80\%$ of the total oscillator strength of  H@$U_{\rm c}$ belongs to a single transition ${1\rm s} \rightarrow {2\rm p}$ - a bizarre property of
the atom under penetrable confinement. Note, earlier \cite{Laughlin,Stevanovic}, the anomaly large value of $f_{1\rm s \rightarrow 2p} \approx 1$ of hydrogen was predicted upon its confinement inside of an impenetrable well ($U_{0}=\infty$, $D =\infty$) of a small radius $r_{0}$. That result, however, is not surprising, since the impenetrable confinement \textit{does} actually compress the atom by driving the $2\rm p$ orbital closer to the $1\rm s$ one.

Looking at figure~$4$ ($r_{0}=1.59$), one meets another set of unusual results. Indeed, one can see that, due to atomic swelling, the excited  $P_{2\rm p}$ and $P_{3\rm p}$ orbitals of H@$U_{\rm c}$ are, first,
pushed far away from the origin, whereas the ground-state function  $P_{1\rm s}$ peaks compactly near $r \approx 1$, as in the free atom ($1\rm s$ atomic swelling does not occur at $r_{0}=1.59$). Thus, second, mostly important, both $P_{2\rm p} \approx 0$ and $P_{3\rm p} \approx 0$
everywhere where $P_{1\rm s} \neq 0$ in H@$U_{\rm c}$,  in contrast to the behaviour of $P_{1\rm s}$, $P_{2\rm p}$ and $P_{3\rm p}$ in the free atom. Correspondingly, the overlap between $P_{1\rm s}$ and $P_{n\rm p}$
functions is practically zero, so that $f_{{1\rm s} \rightarrow {n\rm p}} = 0$, in the compressed atom. This implies, that the compressed atom has lost its $1\rm s \rightarrow n'\ell\pm 1$ discrete photoabsorption spectrum. Hence, the only possibility for the
H($1\rm s$)@$U_{c}$ atom to absorb a photon is exclusively through its photoionization. Thus, the spectra of the free hydrogen, hydrogen under confinement with $r_{0}=1.59$ and hydrogen under confinement with $r_{0}=1.45$
are distinctly different from each other, both qualitatively and quantitatively.

In conclusion, the discussion in the present paper has dealt with the atomic structure and photo-spectra of  hydrogen  under  compression by a repulsive penetrable spherical potential $U_{\rm c}(r)$ of a certain height $U_{0}$,
thickness $\Delta$ and inner radius $r_{0}$. The tendency of sudden atomic swelling rather than contraction upon increasing pressure has been discovered. Profoundly distinct impacts
of atomic swelling on the photo-spectra of hydrogen under confinement have been demonstrated. Specifically, it has been unraveled that atomic swelling can result either in the loss of a discrete photoabsorption spectrum of the atom,
or, on the contrary, in its significant gain, depending on pressure (i.e., the confinement radius $r_{0}$). The findings have been exemplified using of arbitrarily chosen values of $U_{0}$ and $\Delta$. However,  some critical values of $U_{0}$ and $\Delta$ below which the effects will vanish are anticipated. This will happen when $U_{0}$ and $\Delta$ are such that
the confining potential fails to push atomic levels up to the degree needed for the electron to jump from a narrow inner binding well of the effective potential $U_{1\rm s}^{\rm eff}(r)$ into its outer binding well
(see figure $2$), at any $r_{0}$. Other than that, the effects discovered must persist for a broad range of $U_{0}$ and $\Delta$ values. As a follow-up study, it would be interesting to learn how atomic swelling develops in a multielectron atom, where not just one electron but two or more might jump from the inner well into the outer well of $U^{\rm eff}(r)$ at a certain critical value of $r_{0}$,  how the effect could affect electron correlation in the atom, as well as
how all this could modify the interaction of radiation with such a compressed multielectron atom relative to the free atom. The authors are currently working on these topics. As for the present paper, its sole aim has been to demonstrate the existence and importance of  atomic swelling itself, as the first step  towards the understanding of what might happen in atoms confined by repulsive spherical potentials of finite thickness.

\ack
The authors are thankful to Dr. M. Ya. Amusia and Dr. S. T. Manson for careful reading of the manuscript and critical comments. This work was supported by NSF Grant No.\ PHY-$0969386$.

\section*{References}


\begin{thebibliography}{}
%
\bibitem{Michels} Michels A, de Boer J and Bijl A 1937 \textit{Physica} \textbf{4} 981
%
\bibitem{Sommerfeld} Sommerfeld A and Welker H 1938 {\it Ann. Phys.} {\bf 32} 56
%
\bibitem{Jaskolski}  Jask\'{o}lski W 1996 {\it Phys. Rep.} {\bf 271} 1
%
\bibitem{Buchachenko}  Buchachenko A L \textit{ J. Phys. Chem.} B 2001 \textbf{105} 5839
%
\bibitem{RPC'04} Dolmatov V K, Connerade J-P, Baltenkov A S and Manson S T 2004 \textit{Radiat. Phys. Chem.} \textbf{70} 417
%
\bibitem{AQC57} \textit{Theory of Confined Quantum
Sytems: Part One}, edited by  Sabin J R,  Br\"{a}ndas E and Cruz C A, Advances in
Quantum Chemistry (Academic Press, New York, 2009), Vol. 57, pp 334.
%
\bibitem{AQC58} \textit{Theory of Confined Quantum
Sytems: Part Two}, edited by  Sabin J R,  Br\"{a}ndas E and Cruz C A, Advances in
Quantum Chemistry (Academic Press, New York, 2009), Vol. 58, pp 297.
%
\bibitem{ConDolMan00} Connerade J-P, Dolmatov V K and Manson S T 2000 \jpb \textbf{33} L275
%
\bibitem{ConDolMan99} Connerade J-P, Dolmatov V K and Manson S T 1999 \jpb \textbf{32} L395
%
\bibitem{Connerade_book} Connerade J-P 1998 \textit{Highly Excited Atoms} (Cambrigde University Press: Cambridge)
%
\bibitem{ConDol} Connerade J-P and Dolmatov V K 1998 \jpb \textbf{31} 3557
%
\bibitem{ConDolLaksh} Connerade J-P, Dolmatov V K and Lakshmi P A 2000 \jpb \textbf{33} 251
%
\bibitem{ConMansf} Connerade J-P and Mansfield M W D 1975 \textit{Proc. R. Soc.} A \textbf{346} 565
%
\bibitem{Laughlin} Laughlin C 2004 \jpb \textbf{37} 4085
%
\bibitem{Stevanovic} Stevanovi\'{c} L  2010 \jpb \textbf{43} 165002
\end{thebibliography}
\end{document}